\documentclass[letter]{aa}     

\usepackage{graphicx}
\usepackage{txfonts}
\usepackage{lipsum}
\usepackage{subcaption}         
\usepackage{lscape}             
\usepackage{placeins}           

\usepackage{natbib}

\newcommand{\hi} {H {\sc i}}
\newcommand{\hei} {He {\sc i}}
\newcommand{\oi} {O {\sc i}}
\newcommand{\cii} {C {\sc ii}}
\newcommand{\mgii} {Mg {\sc ii}}

\newcommand{\eminus} {e$^-$}
\newcommand{\htwo}{H$_2$}
\newcommand{\htwoo}{H$_2$O}
\newcommand{\hp}{H$^+$}
\newcommand{\hep}{He$^+$}
\newcommand{\hepp}{He$^{2+}$}

\newcommand{\htwop}{H$_2^+$}
\newcommand{\hthreep}{H$_3^+$}

\newcommand{\lalpha}{Lyman-$\alpha$}

\begin{document}

   \title{Vibrationally excited {\htwo} mutes the {\hei} triplet line at 1.08 $\mu$m \\
   on warm exo-Neptunes}


   \author{A. Garc\'ia Mu\~noz\inst{1}
        \and D. De Fazio\inst{2}
        \and D. J. Wilson\inst{3}
        \and K. France\inst{3,4}
        }

   \institute{Universit\'e Paris-Saclay, Universit\'e Paris Cit\'e, CEA, CNRS, AIM, Gif-sur-Yvette, 91191, France\\
             \email{antonio.garciamunoz@cea.fr}
            \and Consiglio Nazionale delle Ricerche, Istituto di Struttura della Materia, Rome, Italy
            \and Laboratory for Atmospheric and Space Physics, University of Colorado, 600 UCB, Boulder, 80309, CO, USA
            \and Department of Astrophysical and Planetary Sciences, University of Colorado, 600 UCB, Boulder, 80309, CO, USA            
}

   \date{Received September 30, 20XX}

 
  \abstract
   {Neptune-sized exoplanets or exo-Neptunes 
are fundamental in the description of exoplanet diversity. Their evolution is sculpted by atmospheric escape,
often traced by absorption in the {\hi} {\lalpha} line at 1,216 {\AA}
and the {\hei} triplet line at 1.08 $\mu$m.
On warm exo-Neptunes HAT-P-11 b, GJ 3470 b and GJ 436 b,
{\hi} {\lalpha} absorption causes 
extreme in-transit obscuration of their host stars.
This suggests that {\hei} triplet line absorption will also be strong, yet it has only been identified on two of these planets.}
{We explore previously unaccounted for processes that might attenuate the 
{\hei} triplet line on warm exo-Neptunes.
In particular,
we assess the role of vibrationally excited {\htwo} to remove the {\hep} ion that acts as precursor of the absorbing He(2$^3S$).
} 
{We formed thermal rate coefficients for this chemical process, leveraging the 
available theoretical and experimental data. The process becomes notably fast at the temperatures expected in the atmospheric layers probed by the 
{\hei} triplet line.
}
{Our simulations show that the proposed process 
severely mutes
the line on GJ 3470 b and causes the nondetection on GJ 436 b. 
The overall efficiency of this mechanism is connected to where
in the atmosphere the {\htwo}-to-H transition occurs and, ultimately, to the amount of high-energy radiation received by the planet.
{The process will be more significant on small exoplanets than on hotter or more massive ones, as for the latter the {\htwo}-to-H transition generally occurs deeper in the atmosphere. }
}
   {Weak {\hei} triplet line
absorption need not imply the lack of a primordial, {\htwo}-He-dominated atmosphere, an idea to bear in mind when interpreting the observations of other small exoplanets.}


   \maketitle


\section{Introduction}
Exo-Neptunes offer precious insight into the transition between the gas giants of total mass dominated by {\htwo}-He and the ubiquitous sub-Neptunes of uncertain composition \citep{venturinietal2020,beanetal2021,ikomakobayashi2025}. 
Although extensively investigated  both observationally and theoretically, 
there remain major gaps in the understanding of exo-Neptunes' nature. 
Some of these gaps are connected to the fact that exo-Neptunes' interiors can often be explained by a range of possibilities in which the outer envelopes are dominated by hydrogen, astrophysical ices (e.g. {\htwoo}) and a combination thereof \citep{nettelmannetal2010,otegietal2020}. 
Infrared (IR) spectroscopy of their atmospheres does not always break this degeneracy. Indeed, a non-negligible number of exo-Neptunes exhibit featureless transmission spectra due to the occurrence of high-altitude clouds and small atmospheric scale heights \citep{knutsonetal2014,grasseretal2024,sunetal2024}.

For exoplanets orbiting close to their host stars, the main mechanism through which they lose mass and evolve is atmospheric escape to space. 
Many escaping atmospheres have been probed by means of absorption lines of {\hi} and {\hei}, 
and of lines of metals such as {\oi}, {\cii} or {\mgii} 
\citep{vidal-madjaretal2003,vidalmadjaretal2004,fossatietal2010,
spakeetal2018,yanhenning2018,garciamunozetal2021,czeslaetal2022,loydetal2025}.
The {\hi} {\lalpha} line at 1,216 {\AA} and the 
{\hei} triplet line at 1.08 $\mu$m are both prominent on 
exoplanets of very disparate conditions and thus useful for comparative studies. The first line arises in absorption from the  ground to the lowermost excited state of the H atom. The
second one arises in
He(2$^3S$){$+$}$h\nu${$\rightarrow$}He(2$^3P$). 
The He(2$^3S$) triplet state is metastable  and, in the layers probed by transmission spectroscopy, produced by radiative recombination, reaction R$_1$: {\hep}{$+$}{\eminus}{$\rightarrow$}He($i$){$+$}{$h\nu$}, and subsequent relaxation of the nascent states \citep{oklopcichirata2018}. 
As the precursor of He(2$^3S$), 
the {\hep} ion partly controls the
absorption line's strength.
The {\hep} ion is mostly produced by
photoionization. 
Its loss is affected
by multiple chemical-collisional-radiative disequilibrium processes \citep{oklopcic2019,garciamunoz2025}, among which reaction R$_1$ is usually important.
Any investigation that builds upon the {\hei} triplet line for characterizing the atmosphere must rely on sophisticated models and on the implementation in them of the leading disequilibrium processes \citep{oklopcichirata2018,
dossantosetal2022,lamponetal2023}. 
Put differently, any findings based on such models will only be as robust as the models are complete and accurate.

To date, the {\hei} triplet line has been detected on $\sim$20 exoplanets, mostly gas giants but also some exo-Neptunes
and one planet 
(GJ 3090 b, optical radius $R_{\rm{p}}$ {$\sim$}2.1$R_{\oplus}$) at the frontier between sub-Neptunes and super-Earths 
\citep{dossantosetal2022,fossatietal2022,vissapragadaetal2022, fossatietal2023,orell-miqueletal2024,guilluyetal2024,massonetal2024,  ahreretal2025}. 
Surprisingly, the number of 
attempted but unsuccessful detections 
is comparable, and includes planets for which past models predicted strong absorption \citep{kasperetal2020,rumenskikhetal2023}. 
The nondetections suggest that there is more to line formation than simply competition between photoionization of ground-state He atoms by Extreme Ultraviolet (XUV; wavelengths {$<$}912 {\AA}), which favours reaction R$_1$,  and of He(2$^3S$) by longer-wavelength photons, which removes it, an overall mechanism that predicts K dwarfs as the best host stars for {\hei} triplet line searches \citep{oklopcic2019}. 
To advance the understanding of the {\hei} triplet line on small exoplanets, we focus on three warm exo-Neptunes, namely:  HAT-P-11 b, GJ 3470 b and GJ 436 b. 
Our uniform treatment of the host stars' spectra aims to minimize potential biases in the results introduced by arbitrarily combining various stellar spectrum sources.

\section{Atmospheric escape from warm exo-Neptunes}
HAT-P-11 b \citep[$R_{\rm{p}}${$\sim$}4.9$R_{\oplus}$, mass $M_{\rm{p}}${$\sim$}25$M_{\oplus}$, 
equilibrium temperature $T_{\rm{eq}}${$\sim$}847 K; ][]{basilicata2024}, 
GJ 3470 b 
\citep[3.9$R_{\oplus}$,
12.6$M_{\oplus}$, 
615 K; ][]{kosiareketal2019}, 
and GJ 436 b 
\citep[4.2$R_{\oplus}$,
23.1$M_{\oplus}$, 
686 K; ][]{turneretal2016} 
are the most extensively investigated exo-Neptunes. They follow nearly polar orbits, 
a feature common to other exo-Neptunes. 
GJ 436 b’s atmospheric metallicity remains  unconstrained because the measured transmission and emission spectra are featureless
\citep{knutsonetal2014,grasseretal2024,mukherjeeetal2025}. 
There are significant uncertainties in the metallicities 
of the other two planets. 
Early observations suggested modest values  \citep{bennekeetal2019,chachanetal2019,sunetal2024}, but the view may be changing and recent JWST data suggest that GJ 3470 b’s atmospheric metallicity is {$\times$}100 solar \citep{beattyetal2024}.

\begin{figure*}
   \centering
   \includegraphics[width=13cm]{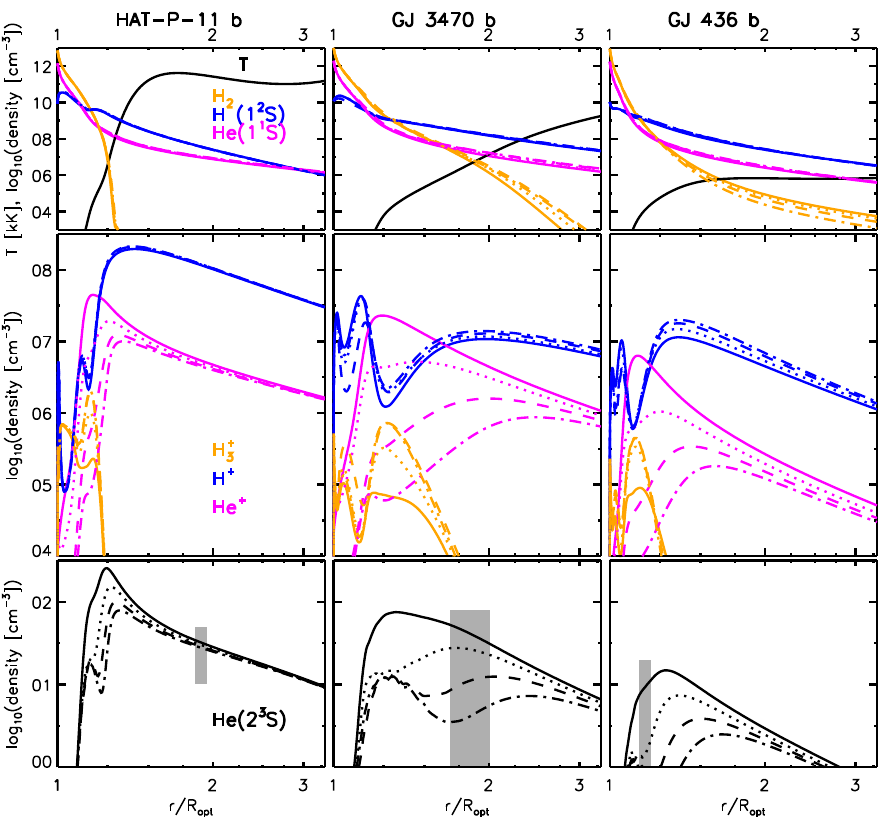}\\
      \caption{\label{bigpanel_fig} 
Profiles of temperature and  neutral densities (top), ion densities (middle) and the 
He(2$^3S$) metastable state density (bottom) for the three warm exo-Neptunes. 
Solid, dotted, dashed and dotted-dashed curves refer to calculations based on 
$k_{{{\rm{R_2}},v=0}}$, $k_{\rm{R_2, LTE}}^{\rm{the}}$, 
$k_{\rm{R_2, LTE}}^{\rm{exp,min}}$ and 
$k_{\rm{R_2, LTE}}^{\rm{exp,max}}$, respectively.
The gray boxes in the bottom panel bracket the range of opaque radii calculated here (Table \ref{req_table}). 
}
\end{figure*}
 
The {\hi} {\lalpha} line has been detected on all three warm 
exo-Neptunes, consistently showing strong absorption that extends well out of the optical transit
\citep{kulowetal2014,ehrenreichetal2015,bourrieretal2018,ben-jaffeletal2022}. 
In the line wings, at hundred or more km/s from the core typically hidden by interstellar medium absorption, the H atoms that enshroud the planets 
obscure their host stars by up to a few times their optical radii \citep{garciamunozetal2020}. 
The {\hei} triplet line has been detected {at each published attempt}
on HAT-P-11 b
\citep{allartetal2018,mansfieldetal2018,allartetal2023,guilluyetal2024}.
The planet's opaque radius in the line core 
is about twice the optical radius. 
Transmission spectroscopy of the 
{\hei} triplet line on GJ 3470 b has resulted in 
a few clear detections, but also 
some nondetections \citep{ninanetal2020,palleetal2020,
allartetal2023,
guilluyetal2024,
massonetal2024}. 
The variability might be caused by temporal changes in the radiative (and possibly corpuscular) output of the host star, a scenario that needs validation by, for example, monitoring the star at X-ray wavelengths over multiple years. 
Unlike for HAT-P-11 b and GJ 3470 b, 
all past searches of the {\hei} triplet line on GJ 436 b have been unsuccessful \citep{nortmannetal2018,
guilluyetal2024,massonetal2024}.
A proposed explanation is that the ratio He$_*$/H$_*$ of helium to hydrogen nuclei on this planet is notably subsolar \citep{rumenskikhetal2023}. 
Naturally, the idea prompts further questions regarding the fate of the helium accreted during planet formation.
Alternatively, it has been shown that the line weakness 
is partly explained by accepting that the host star emits in the XUV less than assumed in the early models \citep{garciamunoz2025,sanz-forcadaetal2025}.
Disequilibrium chemistry in the atmosphere involving vibrationally excited {\htwo} offers an additional viable explanation.

We simulated the escaping atmospheres of the three warm exo-Neptunes with our own model \citep{garciamunoz2025}, generally assuming a solar He$_*$/H$_*$.  
 Each of the planets receives a notably different amount of XUV radiation (Table \ref{SED_table}), with GJ 3470 b and GJ 436 b receiving about 22{\%} and 2.3{\%} of the incident XUV flux on HAT-P-11 b.
The simulations show (Fig. \ref{bigpanel_fig}, top) that the 
transition from {\htwo} (the dominant hydrogen form in the deeper atmospheric layers) into H  
becomes nearly complete at $r$/$R_{\rm{p}}${$\sim$}1.3 for HAT-P-11, but that 
{\htwo} survives to much farther distances on GJ 3470 b and, especially, on GJ 436 b. 
Similar behaviors have been described by past models 
\citep{loydetal2017,ben-jaffeletal2022,garciamunoz2025}, and indicate that the stronger the incident XUV flux the deeper the {\htwo}-to-H transition occurs.
The transformation is mostly driven by ion-neutral 
chemical reactions, ultimately connected to photoionization, and by {\htwo} thermal dissociation where temperatures are high. 
Direct photodissociation of {\htwo} into neutral fragments typically contributes in a minor way to the {\htwo}-to-H transition.

In our original model, the removal of the precursor ion {\hep} in the layers probed by transmission spectroscopy of the 
{\hei} triplet line is controlled by reaction R$_1$.
That model implicitly assumes that {\htwo} is in the vibrational ground state and that the rate coefficient for reaction 
R$_2$: {\hep}{$+$}{\htwo}($v$){$\rightarrow$}He$+${\hp}+H
($v$ is the quantum number; exothermic by 6.5 eV) takes the value 
$k_{{{\rm{R_2}},v=0}}$=3$\times$10$^{-14}$ cm$^3$s$^{-1}$
measured in the laboratory at low temperature \citep{schaueretal1989}
and recommended in astrochemistry applications \citep{mcelroyetal2013}. 
This is five orders of magnitude slower than the collisional limit at which many other exothermic ion-neutral reactions proceed \citep{jonesetal1986}.

\section{Disequilibrium chemistry driven by {\htwo}($v${$>$0})}
Interestingly, there is both experimental and theoretical evidence 
 that reaction R$_2$ becomes fast when {\htwo} 
 is vibrationally excited and that it approaches the collisional limit for $v${$\ge$}2 
 \citep{prestonetal1978,
johnsenetal1980,jonesetal1980,
jonesetal1986,
aguillon1998,defazioetal2019}. 
 To our knowledge, 
the possibility that reaction R$_2$ may control the {\hep} abundance in exoplanet atmospheres and in turn that of He(2$^3S$) has thus far been overlooked. 
Related ideas are however well established in the modelling of photodissociation regions and protoplanetary disks \citep{agundezetal2010,goicoechearoncero2022}.
We compiled the available data on the $v$-resolved rate coefficients for reaction R$_2$ (Appendix \ref{methods_sec}), and formed thermal
rate coefficients $k_{\rm{R_2, LTE}}$ under three scenarios motivated by the origin and uncertainties of the chemical data 
(and satisfying   
$k_{\rm{R_2, LTE}}^{\rm{the}}${$<$}$k_{\rm{R_2, LTE}}^{\rm{exp,min}}${$\leq$}$k_{\rm{R_2, LTE}}^{\rm{exp,max}}$
). $k_{\rm{R_2, LTE}}$ exceeds  $k_{{{\rm{R_2}},v=0}}$ by up to four orders of magnitude in the conditions predicted for the three warm exo-Neptunes (Fig. \ref{kR2LTE_maintext_fig}).
Foreseeably, reaction R$_2$ will take over R$_1$ as a sink for {\hep} where the [{\eminus}]/[{\htwo}] density ratio is  small and the temperatures are high.

\begin{table*}
\caption{Transmission depths. For each bibliographic source/planet, left and right-hand side entries refer to the Excess Absorption (EA) and the planet's opaque radius at the
{\hei} triplet line core \citep{guilluyetal2024}, respectively. 
Last row, mass loss rate calculated here (Methods) for the specified setting. 
Notes. $\dagger$: Opaque radius inferred by us from the EA quoted in the reference, using the planet and star sizes adopted here; 
$\ddagger$: Estimated from their Fig. 5;  
$\triangle$: Measured on a 0.75-{\AA} passband.
}
\label{req_table}
\centering                        
\begin{small}
\begin{tabular}{c c c c}

\hline                 

\hline

Source & HAT-P-11 b & GJ 3470 b & GJ 436 b \\

\hline

\underline{Measurements} & & & \\

\citet{nortmannetal2018}{$\dagger$}  &  & & $<$0.41{\%} ; $<$1.22 \\

\citet{ninanetal2020}{$\ddagger$}  &  & 1.5{\%} ; 1.98  \\
\citet{palleetal2020}{$\dagger$}  &  & 1.50{$\pm$}0.3{\%} ; 1.98{$\pm$}0.15  \\

\citet{allartetal2023}{$\dagger,\triangle$}  & 0.76{$\pm$}0.07{\%} ; 1.75{$\pm$}0.05 & $<$0.64{\%} ; $<$1.50 & \\
\citet{guilluyetal2024}{$\dagger$} & 1.36{$\pm$}0.17{\%} ; 2.17{$\pm$}0.11 & 1.75{$\pm$}0.36{\%} ; 2.10{$\pm$}0.17 & $<$0.42{\%} ; $<$1.22  \\

\citet{massonetal2024}{$\dagger$} & 1.2{$\pm$}0.2{\%} ; 2.1{$\pm$}0.1 & $<$0.9{\%} ; $<$1.7 & $<$0.3{\%} ; $<$1.2 \\

\hline

\underline{Our calculations} & & & \\

${k_{{{\rm{R_2}},v=0}}}$ &1.03{\%} ; 1.95 & 1.55{\%} ; 2.01 &  0.39{\%} ; 1.20\\

$k^{\rm{the}}_{{{\rm{R_2,LTE}}}}$ &0.97{\%} ; 1.91 & 1.27{\%} ; 1.87 & 0.32{\%} ; 1.17 \\

$k^{\rm{exp,min}}_{{{\rm{R_2,LTE}}}}$ &0.92{\%} ; 1.87 & 1.08{\%} ; 1.77 & 0.28{\%} ; 1.15 \\

$k^{\rm{exp,max}}_{{{\rm{R_2,LTE}}}}$ & 0.90{\%} ; 1.85 & 0.93{\%} ; 1.68 & 0.25{\%} ; 1.14 \\

\hline

$\dot{m}$ [g s$^{-1}$] and 
$k^{\rm{exp,min}}_{{{\rm{R_2,LTE}}}}$ & 
2.0$\times$10$^{11}$  & 5.9$\times$10$^{10}$ & 6.4$\times$10$^9$ \\

\hline
\end{tabular}
\end{small}
\end{table*}

We repeated the simulations of the escaping atmospheres using the new $k_{\rm{R_2, LTE}}$.
For the three warm exo-Neptunes, the  {\htwo}, H and He densities change negligibly (Fig. \ref{bigpanel_fig}, top).
In contrast, 
the vibrationally-enhanced 
rates for reaction R$_2$ result in large drops in the {\hep} densities there where {\htwo} remains undissociated 
(Fig. \ref{bigpanel_fig}, middle).  
This occurs simultaneously with the charge transfer 
from {\hep} to {\hp} through reaction R$_2$, and from {\hp} to {\hthreep} through {\hp}+{\htwo}{$\rightarrow$}{\htwop}+H and {\htwop}+{\htwo}{$\rightarrow$}{\hthreep}+H. Reduced 
densities of the precursor ion {\hep} lead to reduced densities of the He(2$^3S$) metastable state (Fig. \ref{bigpanel_fig}, bottom). 
The effect is the largest for GJ 3470 b and GJ 436 b. 
The reason can be traced to the overlap on these planets of the layer where {\htwo} remains undissociated and the layer where {\hep} remains abundant. 
Our simulations reveal the importance of the {\htwo}-to-H transition as a factor controlling the He(2$^3S$) density. The {\htwo} survival to far distances on both GJ 3470 b and GJ 436 b opens the possibility, unexplored here, that the {\htwo} vibrationally excited states could be populated by stellar photoexcitation or by collisions with thermal electrons and become detectable in absorption in the Werner and Lyman bands at Far-UV (FUV) wavelengths \citep{morganetal2022}. 

Transmission spectroscopy is sensitive to the outermost atmospheric layers.
We generated spectra of the {\hei} triplet line based on the above simulations, and extracted from each spectrum two transmission depth indicators (the excess absorption and the planet's opaque radius, both specified at line core \citep[][Table \ref{req_table}]{guilluyetal2024,massonetal2024}. 
For HAT-P-11 b, the transmission depths depend weakly on the  rate coefficient for reaction R$_2$, and are consistent with the  measurements. 
As expected,  
the effect of the vibrationally-enhanced reaction R$_2$ 
on the transmission depths is significant for GJ 3470 b and GJ 436 b. 
The simulations adopting the thermal rate coefficients bring the transmission depths in closer agreement with the occasional detections on GJ 3470 b and the systematic nondetections on GJ 436 b. 
These simulations confirm that the He($2^3S$) density on these planets is significantly affected by disequilibrium chemistry driven by vibrationally excited {\htwo} through reaction R$_2$.

For the standard XUV flux of 278 erg cm$^{-2}$s$^{-1}$ incident on GJ 436 b (Table \ref{SED_table}), the most recent upper limit on the excess absorption in the {\hei} triplet line (0.3{\%} \citep{massonetal2024}; Table \ref{req_table})
is explained by He$_*$/H$_*$$<${0.06}, $<$0.08, 
$<$0.10 or  $<$0.12, 
if $k_{{{{\rm{R_2}},v=0}}}$, $k^{\rm{the}}_{{{\rm{R_2,LTE}}}}$, 
$k^{\rm{exp,min}}_{{{\rm{R_2,LTE}}}}$ or $k^{\rm{exp,max}}_{{{\rm{R_2,LTE}}}}$, are adopted in the model, respectively. 
Clearly, the choice of the rate coefficient for reaction R$_2$ limits the information that can potentially be inferred from {\hei} triplet line measurements. New determinations, experimental or theoretical, of the rate coefficient for reaction R$_2$ over a broad range of temperatures are needed.

\section{Discussion and outlook}
Conceivably, the {\hei} triplet line might be used to assess whether an atmosphere is {\htwo}-He-dominated
and therefore primordial or, 
alternatively, metal-rich and secondary. 
The idea is appealing, especially in its application to small exoplanets for which 
IR molecular spectroscopy is challenging but for which the 
large spatial scales associated with the escaping atmosphere 
produce detectable atomic signatures \citep{garciamunozetal2020,garciamunozetal2021,ahreretal2025}.
Without attempting to explore the full range of possibilities here, we simulated the atmosphere of a virtual sub-Neptune
($R_{\rm{p}}${$\sim$}2.2$R_{\oplus}$;   
 same bulk density as GJ 436 b, about half its gravity)
orbiting GJ 436 at the same distance as GJ 436 b. 
The calculations (Fig. \ref{bigpanel_subgj436b_fig}) show that the {\hei} triplet line is notably weaker than for GJ 436 b and that {\htwo} remains undissociated to farther distances. The {\htwo} survival to high altitudes is very detrimental to the strength of the 
{\hei} triplet line, for which we obtain excess absorptions between 0.08{\%} (for $k_{{{{\rm{R_2}},v=0}}}$) and 0.03{\%}
(for $k^{\rm{exp,max}}_{{{\rm{R_2,LTE}}}}$).
On the positive side, the {\htwo} survival 
 suggests that vibrationally excited {\htwo} might become detectable at FUV wavelengths. 
Overall, the large drop in the predicted {\hei} triplet line strength when going from 
$k_{{{{\rm{R_2}},v=0}}}$ to 
$k^{\rm{exp,max}}_{{{\rm{R_2,LTE}}}}$ cautions against simplified interpretations in which absence of {\hei} triplet line absorption might be taken as evidence against a {\htwo}-He-dominated atmosphere.

{
The atmospheres of warm exo-Neptunes are fundamentally different to those of hotter {or heavier} exoplanets (\textit{e.g.} hot Jupiters, for which there are many {\hei} triplet line detections). On the latter, the {\htwo}-to-H transition generally occurs deep in the atmosphere, which makes reaction R$_2$ ineffective at controlling the 
line strength. 
}
{In contrast, }
our work shows that the {\htwo} survival {to far distances on small exoplanets} significantly affects the {\hei} triplet line that is often utilized as a tracer of atmospheric escape. 
The continuing discovery and characterization of other warm exo-Neptunes will provide additional opportunities to test these ideas, especially if there exists a concerted effort to detect the {\hi} {\lalpha} and {\hei} triplet lines on them and to constrain their host stars' high-energy emission.
Our work reveals the importance of chemistry mediated through vibrational states of molecules, and the need to take such effects into account in future interpretation work.


\begin{acknowledgements}
DDF acknowledges CINECA  (ISCRA initiative) for availability of high performance computing resources and support.      
\end{acknowledgements}

%

\bibliographystyle{aa} 
\bibliography{mybiblio} 


\clearpage

\begin{appendix}




\section{Further discussion}

To explore the broader implications of the vibrationally enhanced reaction R$_2$, we simulated GJ 436 b-like planets (and atmospheres of solar He$_*$/H$_*$)  that receive XUV fluxes in the range from 30 to 20,000 erg cm$^{-2}$s$^{-1}$. 
For reference the incident XUV flux at Earth is of a few erg cm$^{-2}$s$^{-1}$. 
In practice, we scaled the  stellar XUV spectra  without modifying the incident fluxes at longer wavelengths. 
The exercise provides also insight into GJ 436 b
at epochs when its host star might have experienced a different activity or the exoplanet might have been on a different orbit. 
The vibrationally enhanced reaction R$_2$ has a major relative impact  on the transmission depth for low-to-moderate XUV fluxes, but the impact is minor for high XUV fluxes (Fig. \ref{EA_XUV_fig}).
The reason is that high XUV fluxes lead to atmospheres in which the {\htwo}-to-H transition occurs deeper, making the removal of the precursor ion {\hep} 
through reaction R$_2$
progressively marginal. 
Based on these calculations, we speculate that a time-varying radiation environment experienced by the planet due to, for example, an activity cycle, rotational modulation or enhanced flare activity of the host star
might be the causes of the temporal variability on GJ 3470 b. A multi-year transit campaign with contemporaneous X-ray and FUV observations to track the XUV output at the same time as tracking the {\hei} triplet line on this warm exo-Neptune will help test these scenarios that, if proven correct, might serve as an indirect method for monitoring the high-energy radiative output of this and other host stars.

\begin{figure}
   \centering
   \includegraphics[width=15cm]{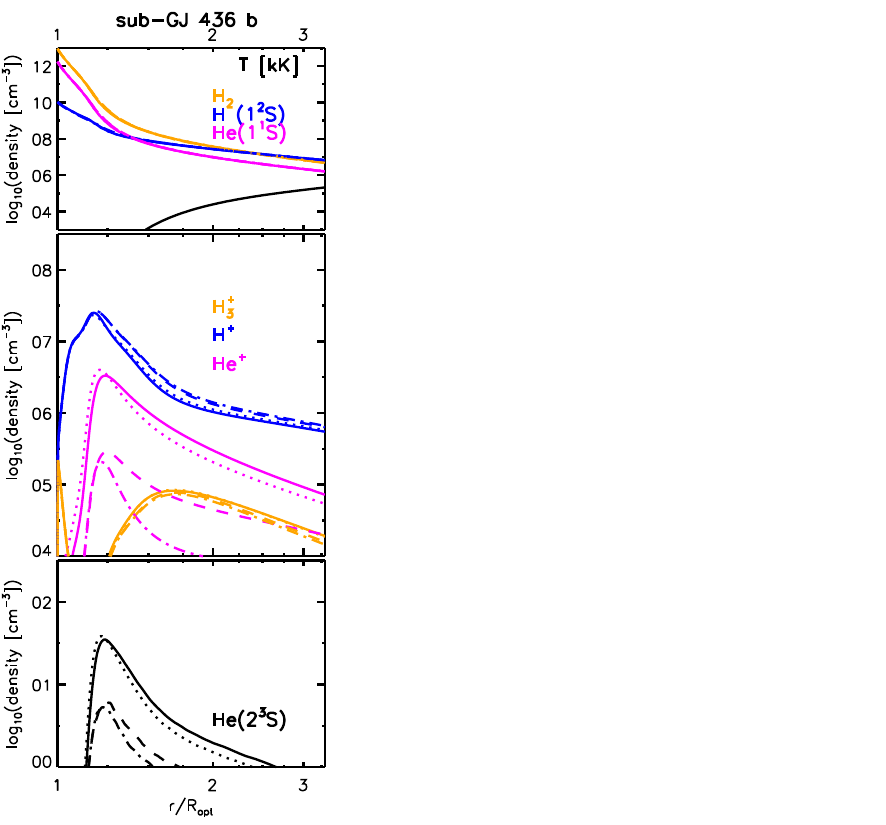}\\
      \caption{\label{bigpanel_subgj436b_fig} 
Same as Fig. \ref{bigpanel_fig}, for the virtual sub-Neptune motivated by GJ 436 b described in the text. 
{The {\hei} triplet line is one of the few atmospheric features detectable on sub-Neptunes with current technology \citep[as in the case of GJ 3090 b, ][]{ahreretal2025}. The limited set of simulations presented in this figure 
confirm that {\htwo} is likely to survive to far distances on them, which has a significant effect on the {\hei} triplet line strength.}
}
\end{figure}
\begin{figure}[h!]
\centering
\includegraphics[width=9cm]{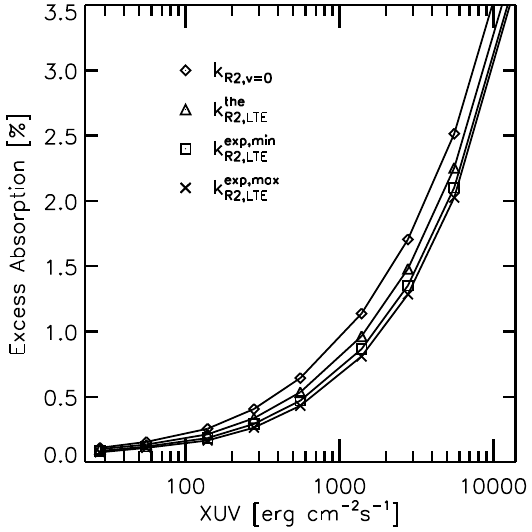}\\
\caption{\label{EA_XUV_fig} 
Excess absorption predicted for a GJ 436 b-like exoplanet under a range of XUV fluxes. On its current orbit, GJ 436 b receives a XUV flux of about 278 erg cm$^{-2}$s$^{-1}$ (Table \ref{SED_table}). Simulations based on the various rate coefficients for reaction R$_2$. 
{For small XUV fluxes, the EA varies by up to a factor of 2 depending on the rate coefficient for reaction R$_2$ that is adopted. For strong XUV fluxes, the choice of reaction rate coefficient is, in relative terms, weaker. This figure complements the EA quoted in Table \ref{req_table} for HAT-P-11 b, GJ 3470 b and GJ 436 b.}
}
\end{figure}

\section{\label{methods_sec}Methods}

\subsection{Rate coefficient for reaction R$_2$}

We formed thermal or LTE rate coefficients for reaction R$_2$, 
$k_{{{\rm{R_2}},\rm{LTE}}}$, from the information available on the $v$-resolved rate coefficients, $v$ referring to the 
vibrational quantum number in {\htwo}($v$).
LTE is used to mean that the {\htwo}($v$) population is assumed to be in local thermodynamic equilibrium, 
described by a truncated Boltzmann distribution of states $v${$\le$2} at the gas kinetic temperature. We calculated the transition probabilities $A_{v'v''}$ for spontaneous emission  $v'$=1--2{$\rightarrow$}$v''${$<$}{$v'$} 
in {\htwo}, finding they are $\lesssim$10$^{-6}$ s$^{-1}$ and therefore much smaller than the  deexcitation rates for collisions with {\htwo} or H 
at the pressures of interest, which gives evidence that the relative 
{\htwo}($v${$\le$2}) populations are dictated by collisions {and thus thermalized}. At the usual temperatures of exoplanet atmospheres, the {\htwo}($v${$>$2}) population is negligible.

We calculate the LTE rate coefficient as:
$$k_{{{\rm{R_2}},\rm{LTE}}}={f_{v=0}}{k_{{{\rm{R_2}},v=0}}}+{f_{v=1}}{k_{{{\rm{R_2}},v=1}}}+{f_{v=2}}{k_{{{\rm{R_2}},v=2}}},$$
where $f_{v}$=${\exp{(-E_v/kT)}}/Z$ is the {\htwo}($v$) relative abundance, and {$k_{{{\rm{R_2}},v}}$} is the $v$-resolved
rate coefficient for collisions of {\hep} with {\htwo}($v$). $Z$=$\sum_{v\le2}{\exp{(-E_{v}/kT)}}$ is the truncated partition function. 
The energies $E_v$ are from a compilation \citep{fantzwunderlich2006}; $k$ and $T$ are Boltzmann's constant and temperature, respectively.
Our calculated $f_{v}$ are consistent with the relative abundances calculated taking also into account the rotational structure of the molecule.

We collected the  information on the {$k_{{{\rm{R_2}},v}}$}
from a variety of sources that include theoretical calculations and laboratory experiments. 
Although there is general consensus in their qualitative behaviors, 
there remain discrepancies in their quantitative values. 
To account for these uncertainties and explore their implications, we formed three LTE rate coefficients, one of them based on theoretical calculations, which we term $k^{\rm{the}}_{{{\rm{R_2}},\rm{LTE}}}$, 
and two of them based on experimental constraints, which we term
$k^{\rm{exp,min}}_{{{\rm{R_2}},\rm{LTE}}}$ and 
$k^{\rm{exp,max}}_{{{\rm{R_2}},\rm{LTE}}}$. Ideally, future work by chemists will solve these discrepancies. 

\begin{figure}[h!]
   \centering
   \includegraphics[width=9cm]{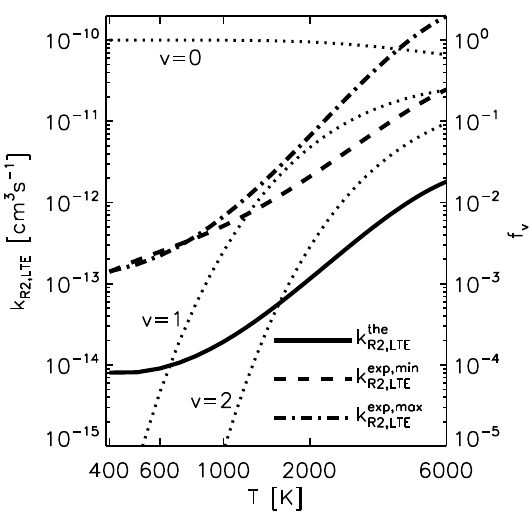}\\
      \caption{\label{kR2LTE_maintext_fig} 
Thermal (Local Thermodynamic Equilibrium, LTE) rate coefficients for reaction R$_2$ implemented in our simulations (see text for details). Also, relative abundances $f_v$ for {\htwo}($v${$\leq$2}) are shown as the dotted lines. 
The experiment-based rate coefficients are consistent with the measurements at temperatures between 400 and 700 K \citep{johnsenetal1980}. 
For reference, 
the low-temperature measurements \citep{schaueretal1989} recommended for astrochemical applications
\citep{mcelroyetal2013}
are $\sim$3$\times$10$^{-14}$ cm$^3$s$^{-1}$.
}
\end{figure}

\subsubsection{Theory-based rate coefficient}

We adopted the {$k^{\rm{the}}_{{{\rm{R_2}},v=0}}$} and {$k^{\rm{the}}_{{{\rm{R_2}},v=1}}$} 
(with both $v$=0 and $v$=1 in their rotational ground states) obtained in quantum dynamical calculations  up to 2,000 K
\citep{defazioetal2019}, and  
assumed they become temperature-independent and equal to 1.30$\times$10$^{-14}$ and 1.39$\times$10$^{-12}$ cm$^3$s$^{-1}$, respectively, for $T${$>$}2,000 K. 
Importantly, test calculations suggest that the
{\htwo}($v$=0) reactivity is very sensitive to rotational excitation, which might enhance the efficiency of reaction R$_2$ as the temperature increases well above the values adopted here, which should therefore be seen as lower limits.
We assumed 
{$k^{\rm{the}}_{{{\rm{R_2}},v=2}}$}=10$\times${$k^{\rm{the}}_{{{\rm{R_2}},v=1}}$} based on available semi-classical cross section calculations at collision energies $\sim$2 eV \citep{aguillon1998}. 
Future rate coefficient calculations should include collisions at energies below 1 eV
with {\htwo}($v${$\le$}2) in a broad range of rotational states.

The parameterization:
$$
\ln{
k^{\rm{the}}_{{{\rm{R_2}},\rm{LTE}}}}=
-1713.16/T + 252.396 $$
$$-110.507  \ln{T} + 14.1998 (\ln{T})^2 -0.596071 (\ln{T})^3
$$
reproduces the theory-based LTE rate coefficient to within 15{\%} from 200 to 10,000 K.

\subsubsection{Experiment-based rate coefficients}

We adopted the available experimental measurements between 400 and 700 K
for the collisions of {\hep} and {\htwo} \citep{johnsenetal1980},
which we assumed to represent 
the combination of 
${f_{v=0}}{k^{\rm{exp}}_{{{\rm{R_2}},v=0}}}+{f_{v=1}}{k^{\rm{exp}}_{{{\rm{R_2}},v=1}}}$.  
The measurements are well described by a power law 
of the type {$\propto$}$T^m$ with $m$=1.4309776, and we used this law 
to represent the contribution of collisions with {\htwo}($v${$\le$}1) over the full range of temperatures. 
To account for the contribution of collisions with {\htwo}($v$=2), 
we added {${f_{v=2}}{k^{\rm{exp}}_{{{\rm{R_2}},v=2}}}$}, where  ${k^{\rm{exp}}_{{{\rm{R_2}},v=2}}}$ was borrowed from published \citep{jonesetal1986} 
lower 
(${k^{\rm{exp,min}}_{{{\rm{R_2}},v=2}}}$=1.8$\times$10$^{-10}$ cm$^3$s$^{-1}$)
and upper (${k^{\rm{exp,max}}_{{{\rm{R_2}},v=2}}}$=
1.8$\times$10$^{-9}$ cm$^3$s$^{-1}$) estimates. 
We formed 
$k^{\rm{exp,min}}_{{{\rm{R_2}},\rm{LTE}}}$ and
$k^{\rm{exp,max}}_{{{\rm{R_2}},\rm{LTE}}}$ by 
using 
${k^{\rm{exp,min}}_{{{\rm{R_2}},v=2}}}$ and
${k^{\rm{exp,max}}_{{{\rm{R_2}},v=2}}}$ respectively. The parameterizations:
$$
\ln{
k^{\rm{exp,min}}_{{{\rm{R_2}},\rm{LTE}}}}=
-4435.40/T + 336.468 $$
$$-131.919  \ln{T} + 15.9440 (\ln{T})^2 -0.636685 (\ln{T})^3
$$
$$
\ln{k^{\rm{exp,max}}_{{{\rm{R_2}},\rm{LTE}}}}=
-3215.06/T + 370.797 $$
$$-152.621  \ln{T} + 19.3836 (\ln{T})^2 -0.807819 (\ln{T})^3
$$
reproduce the experiment-based LTE rate coefficients to within 30{\%} from 400 to 10,000 K.

\begin{figure}
   \centering
   \includegraphics[width=6.5cm]{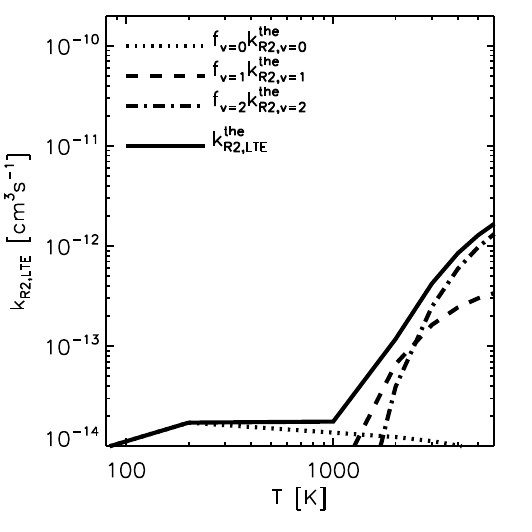}\\
   \vspace{-0.3cm}
   \includegraphics[width=6.5cm]{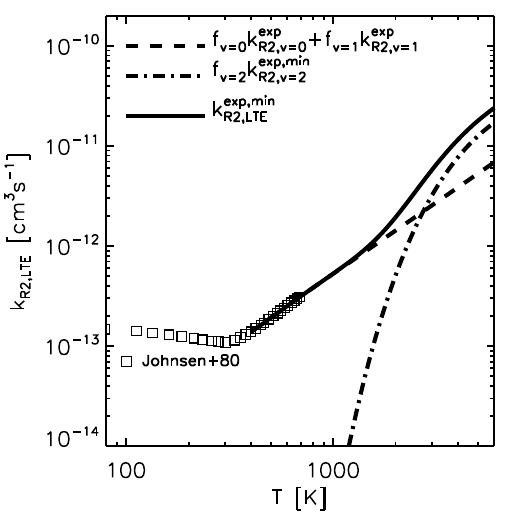}\\
   \vspace{-0.3cm}   
   \includegraphics[width=6.5cm]{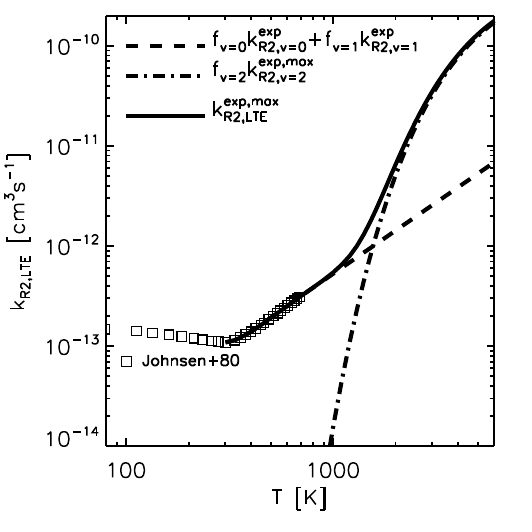}      
      \caption{\label{kR2LTE_fig} 
LTE rate coefficients for reaction R$_2$.
Top: theory-based value. Middle and Bottom: 
Experiment-based values; \citet{johnsenetal1980} measurements are shown for reference.
}
\end{figure}

Figure \ref{kR2LTE_fig} shows the three  $k_{{{\rm{R_2}},\rm{LTE}}}$ 
as a function of temperature.
They differ by up to $\times$100 at the
highest temperature in the plot. The differences call for new constraints either from quantum calculations or experiments. 
This said, both theory and experiments 
have given substantial evidence 
\citep{prestonetal1978,johnsenetal1980,jonesetal1986,schaueretal1989,aguillon1998,defazioetal2019} 
that {$k_{{\rm{R_2}},v=0}$}{$\ll$}{$k_{{\rm{R_2}},v=1}$}{$\ll$}{$k_{{\rm{R_2}},v=2}$}
and this translates into LTE rate coefficients for reaction R$_2$ that vary by 3-4 orders of magnitude between ambient temperature and a few thousand Kelvin. Although exothermic by at least 6.5 eV, reaction R$_2$ exhibits an internal barrier \citep{prestonetal1978}
that makes it unusually slow at ambient temperature. 
Vibrational excitation facilitates the tunneling through the barrier and enhances its reactivity.

\subsection{Stellar SEDs}

Hydrodynamic escape is driven by stellar photons incident on the atmosphere 
that transfer some of their energy to the internal modes of the gas atoms and molecules. The heated atmosphere expands and accelerates into space, setting off a bulk outflow. 
Our simulations show that {\htwo}, H, He and {\hep} are relatively abundant at some altitude and thus prime candidates to drive the gas acceleration. 
The longest wavelength at which these species engage in continuum absorption leading to photofragmentation is {$\sim$}1,200 {\AA}, the threshold for the first step in the sequence 
{\htwo}($X,v$)+{$h\nu$}{$\rightarrow$}{\htwo}($B$, $C$){$\rightarrow$}H+H \citep{stecherwilliams1967}. 
In truth, the threshold depends  on the details of the vibrational population of the electronic ground state 
{\htwo}($X,v$) and thus on temperature. 
If the description of the stellar 
spectral energy distribution
(SED) at $\lambda${$\lesssim$}1,200 {\AA} is key for the outflow modelling, the helium modelling is sensitive to the wavelengths that enable 
He+$h\nu$($\lambda${$<$}504{\AA}){$\rightarrow$}{\hep}+{\eminus}, 
{\hep}+$h\nu$($\lambda${$<$}228{\AA}){$\rightarrow$}{\hepp}+{\eminus} and 
He($2^3S$)+$h\nu$($\lambda${$<$}2,600{\AA}){$\rightarrow$}{\hep}+{\eminus}, as photoionization contributes to the chemical balance of the {\hep} precursor and the He(2$^3S$) metastable.

The high-energy spectra of the host stars of interest in our study
have been studied multiple times \citep{franceetal2016,peacockeal2019,sanz-forcadaetal2025,wilsonetal2025}.
We adopted the HAT-P-11, GJ 3470 and GJ 436 SEDs at $\lambda${$\lesssim$}1,200 {\AA} 
from the X-Exoplanets database \citep{sanz-forcadaetal2025}. 
For the longer wavelengths, we utilized the best available SEDs obtained through the Mega-MUSCLES project \citep{franceetal2016}.
Specifically, for 
HAT-P-11 (stellar type K4, 29 days of rotation period) we adopted data for 
Epsilon Indi (K4-5, 35 days),   
for GJ 3470 (M1.5, 21 days) we adopted data for GJ 649 (M1, 23 days), 
and for GJ 436 (M3.5 V) we adopted the corresponding SED reported in the MUSCLES database.  
Assessing the reliability of the SEDs is 
challenging, partly because the available direct information is limited 
and partly because the stellar radiative output is itself variable over multiple timescales. 
Our approach does at the very least guarantee a uniform treatment for all three host stars.

\begin{figure}
   \centering
   \includegraphics[width=9cm]{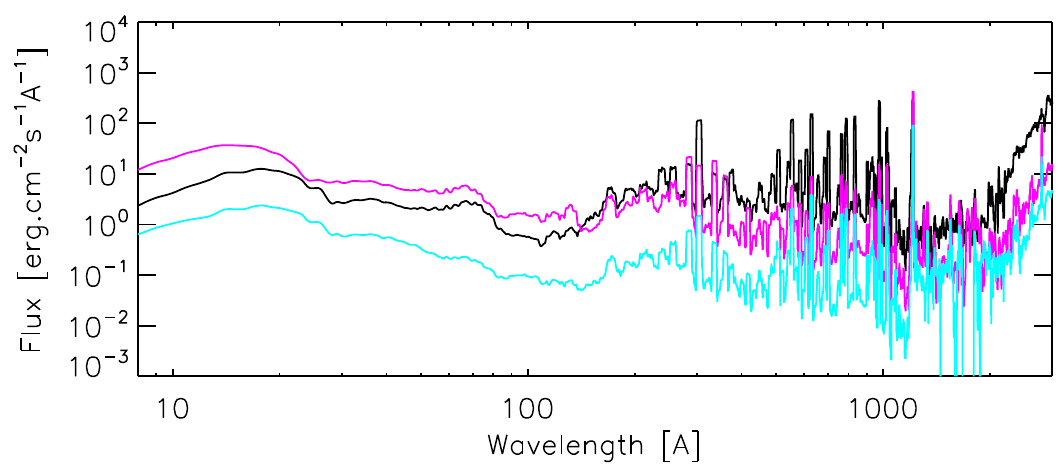}
      \caption{\label{SED_fig} Stellar fluxes incident on the planets, 
      degraded to low resolution for clarity of presentation. 
      Black: HAT-P-11 b; Magenta: GJ 3470 b; Cyan: GJ 436 b.
       }
\end{figure}

Figure \ref{SED_fig} shows the stellar fluxes received by the planets at their orbital positions, and Table \ref{SED_table} summarizes some integrated quantities.

\begin{table*}
\caption{Some integrated properties of the adopted stellar SED at the planet orbits.}
\label{SED_table}
\centering                        
\begin{tabular}{r r r r r r r r}        
\hline                 
\multicolumn{1}{c}{} & \multicolumn{7}{c}{Stellar flux [erg cm$^{-2}$s$^{-1}$] over the specified bands } \\
\multicolumn{1}{c}{Distance} & 0-100 {\AA} &  0-228 {\AA} & 0-504 {\AA} & 0-912 {\AA} & 912-1200 & 1214-1220 {\AA} & 912-2600 {\AA} \\
\hline
\multicolumn{1}{c}{HAT-P-11 b} & 299 & 647  & 3229 & 11907 & 4644 & 2512 & 20590 \\
\multicolumn{1}{c}{GJ 3470  b} & 762 & 1044 & 2029 & 2642  & 461  & 4239 & 5988 \\
\multicolumn{1}{c}{GJ 436 b}   & 53  & 70   & 127  & 278   & 70   & 895  & 1316 \\

\hline
\end{tabular}
\end{table*}

\subsection{Hydrodynamic modelling}

The calculations were performed with a numerical model that solves simultaneously the mass, momentum and energy conservation equations of a gas 
in a one-dimensional spherical-shell atmosphere. 
It includes about 210 chemical-collisional-radiative processes and 20 species of hydrogen and helium plus thermal electrons. {The list of species does not include heavier elements (metals). We will explore the significance of metal-based chemistry in the future.}
The effect of non-thermal electrons produced by ionizing radiation
on the chemistry and the radiative transfer is considered self-consistently \citep{garciamunozbataille2024} without introducing \textit{ad-hoc} efficiencies. 
The model is well suited for investigating the transition from a molecular, {\htwo}-dominated gas to an atomic, mostly ionized plasma. 
\\

The original implementation \citep{garciamunoz2025} is extended to include reaction R$_2$ for an LTE population of {\htwo}. 
We additionally extended the helium chemistry network and the corresponding radiative transfer by adding the double-charge ion {\hepp} (energy $E$({\hepp}){$-$}$E$({\hep})=54.4 eV; {\hep}
remains the only form of the single-charge ion and is assumed to represent the ground electronic state), and the radiative 
{\hep}+{$h\nu$}{$\leftrightarrow$}{\hepp}+{\eminus} 
and charge-exchange
{\hepp}+{H}{$\rightarrow$}{\hep}+{\hp} processes. 
Both the photoionization cross sections from the electronic ground state of {\hep} and the total radiative recombination rate coefficients are borrowed from the NORAD database 
\citep{nahar2010,nahar2020}. For reference, Table \ref{radrecomhep_table} summarizes the radiative recombination rate coefficients at a few temperatures. 
{\hep} photoionization conceivably contributes to the gas opacity at high altitudes and modifies the ion abundance there. 
Our calculations show however that
this affects negligibly the atmosphere where the {\hei} triplet line is formed.
We calculated the charge-exchange rate coefficient directly from the cross sections \citep{westetal1982}. Its value is well approximated 
from 200 to 10,000 K by the temperature-independent value 1.70$\times$10$^{-13}$ cm$^3$s$^{-1}$. This is consistent with the value recommended in a compilation \citep{arnaudrothenflug1985} 
but more than an order of magnitude larger than the value recommended in another compilation \citep{kingdonferland1996}.

The quoted mass loss rates are calculated from $\dot{m}$=4$\pi${$\rho u r^2$}, where  
$\rho$ is the volume density of the gas, 
$u$ is its bulk velocity and $r$ is the radial distance to the center of the planet. We solve the gas flow along the substellar line, without attenuating the incident stellar flux to take into account slanted irradiation near the terminators. This may artificially boost the mass loss rate over its true value by a case-dependent factor of $\sim$2. The actual value can only be determined by means of multi-dimensional calculations, which are beyond the scope of this work. 

\begin{table}
\caption{Total rate coefficient [cm$^3$s$^{-1}$] at selected temperatures for radiative recombination, 
{\hepp}+{\eminus}{$\rightarrow$}{\hep}+$h\nu$, where the end ion includes all excitation states. Also indicated, the wavelength threshold for continuum emission.}
\label{radrecomhep_table}
\centering                        
\begin{tabular}{r c c c c}        
\hline                 
   &    \multicolumn{4}{c}{$T$ [K]}    \\
\multicolumn{1}{c}{[{\AA}]}    & 500 &  2,000     &    5,000     &    10,000    \\
\hline
227.8     & 1.50E$-$11 & 6.27E$-$12 & 3.46E$-$12 & 2.18E$-$12    \\
\hline
\end{tabular}
\end{table}

\subsection{Spectral modelling}

\begin{figure}
   \centering
   \includegraphics[width=10cm]{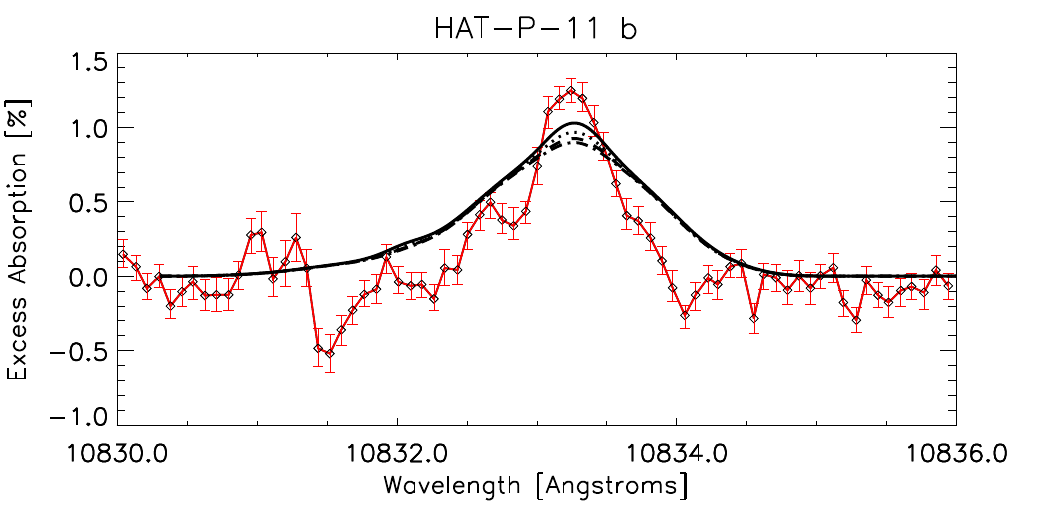}
   \includegraphics[width=10cm]{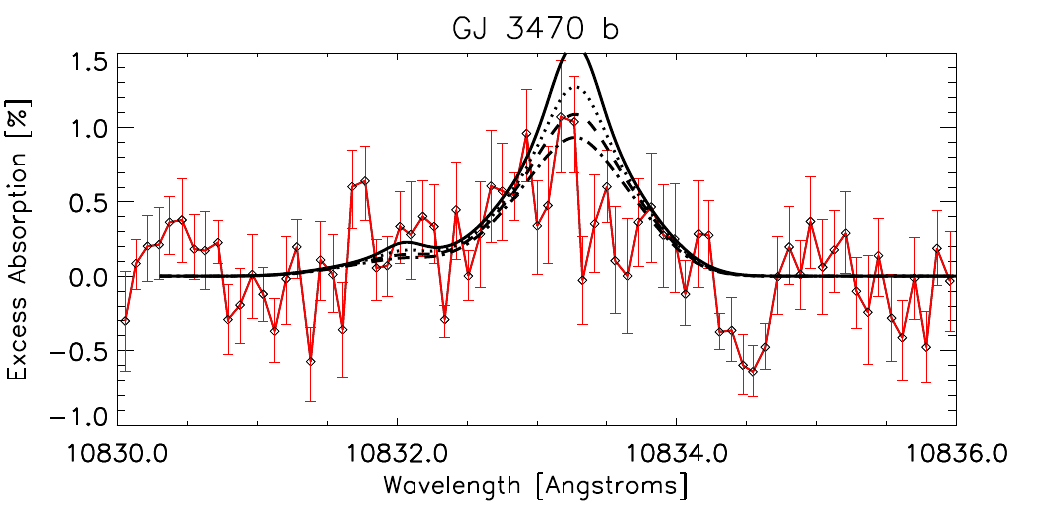}   
   \includegraphics[width=10cm]{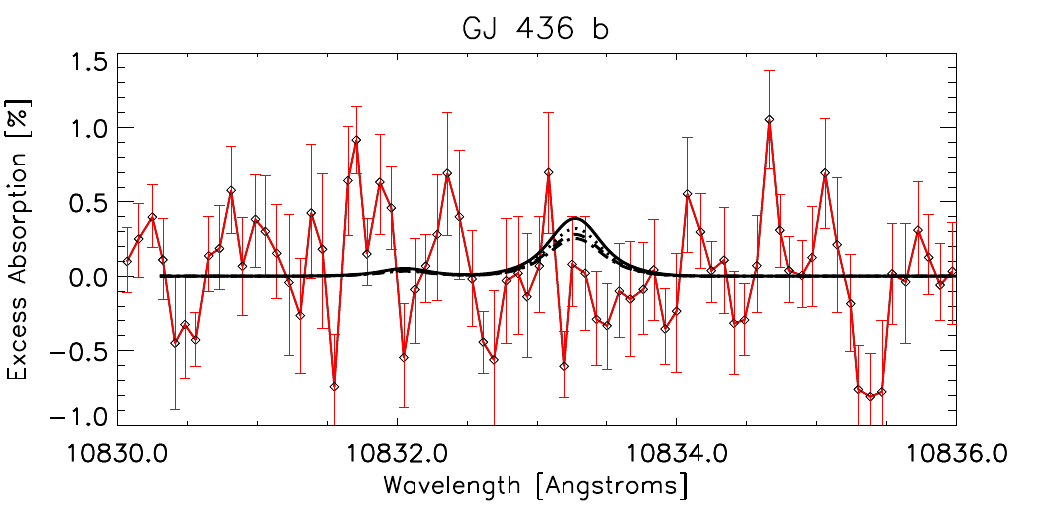}      \caption{\label{transmissionspectra_fig} 
Black: Synthetic spectra  based on the atmospheric models of Fig. \ref{bigpanel_fig} of Main Text (same pattern code). Red: Measured spectra and uncertainties \citep{massonetal2024}.
}
\end{figure}

We have calculated the transmission depths using a previously presented methodology \citep{garciamunoz2025}. The calculation takes into account the Doppler shift introduced by the gas escaping towards and away from the observer in the assumed spherical shell geometry. 
Figure \ref{transmissionspectra_fig} summarizes the transmission spectra for the simulations presented in Fig. \ref{bigpanel_fig}.

%





\FloatBarrier 
\clearpage

\end{appendix}
\end{document}